\documentclass[twocolumn,showpacs,showkeys,a4paper,superscriptaddress]{revtex4} 

\usepackage[english]{babel}
\usepackage{amssymb,amsbsy} 
\usepackage{amsmath}
\usepackage{graphicx}
\usepackage{verbatim}
\usepackage{color}
\usepackage{subfigure}
\usepackage{hyperref}
\usepackage{booktabs,caption}
\usepackage{float}

\newcommand{\eqeqref}[1]{Eq.~\eqref{#1}}

\newcommand{\secref}[1]{Section~\ref{#1}}

\newcommand{\mb}[1]{\mathbf{#1}}

\begin{document}

\title{Non-extensive statistical mechanics of a self-gravitating gas} 

\author{L. F. Escamilla-Herrera}
\email{lenin.escamilla@correo.nucleares.unam.mx} 
\affiliation{Instituto de Ciencias Nucleares, Universidad Nacional 
	Aut\'{o}noma de M\'{e}xico, Mexico City 04510, M\'{e}xico.}

\author{C. Gruber}
\email{christine.gruber@uni-oldenburg.de} 
\affiliation{Hanse-Wissenschaftskolleg Delmenhorst, Germany}
\affiliation{Institut f\"ur Physik, Universit\"at Oldenburg, 
	D-26111 Oldenburg, Germany}

\author{V. Pineda-Reyes}
\email{viridiana.pineda@correo.nucleares.unam.mx} 
\affiliation{Instituto de Ciencias Nucleares, Universidad Nacional 
	Aut\'{o}noma de M\'{e}xico, Mexico City 04510, M\'{e}xico.}

\author{H. Quevedo}
\email{quevedo@nucleares.unam.mx} 
\affiliation{Instituto de Ciencias Nucleares, Universidad Nacional 
	Aut\'{o}noma de M\'{e}xico, Mexico City 04510, M\'{e}xico.}
\affiliation{Dipartimento di Fisica and ICRA, Universit\`a di Roma ``La Sapienza", I-00185 Roma, Italy}
\affiliation{Institute of Experimental and Theoretical Physics, 
Al-Farabi Kazakh National University, Almaty, Kazakhstan}



\date{\today}

\begin{abstract}
The statistical mechanics of a cloud of particles interacting via 
their gravitational potentials is an old problem which encounters 
some issues when the traditional Boltzmann--Gibbs statistics is 
applied. In this article, we consider the generalized statistics 
of Tsallis and analyze the statistical and thermodynamical 
implications for a self-gravitating gas, obtaining analytical and 
convergent expressions for the equation of state and specific heat 
in the canonical as well as microcanonical ensembles. Although our 
results are comparable in both ensembles, it turns out that only in 
the canonical case the thermodynamic quantities depend explicitly 
on the non-extensivity parameter, indicating that the question of 
ensemble equivalence for Tsallis statistics must be further 
reviewed.
\end{abstract}

\keywords{Statistical Mechanics, Self-gravitating systems, Tsallis non-extensive statistics}
\pacs{05.20.Jj, 95.30.Sf, 95.30.Tg}

\maketitle

\section{Introduction}

Statistical mechanics of many-body systems and the connection with 
principles and notions of thermodynamics have been a powerful 
tool in physics for a long time, and have established themselves 
as a very robust framework continuing to hold even in the face of 
big theoretical revolutions and paradigm shifts. However, systems 
where gravity is the dominant interaction have always posed a 
challenge even for thermodynamics, especially since the introduction 
and acceptance of general relativity. The inclusion of gravitational 
forces into statistical models, indeed representing only an example 
for the general problem of long-range forces in physics, has led 
very early on to puzzles or incompatibilities with well-established 
thermodynamic notions~\cite{2009Campa,2003Oppenheim,2001Barre}. The 
Tolman-Ehrenfest effect 
\cite{1930Tolma,1930Tolmb,1930Tolmc} is a very good illustration -- 
considering a compact object made up of a perfect fluid described 
by a spherical metric, it turns out that -- supposing an isolated 
system -- a stationary (non-changing) state can be achieved, but 
that thermodynamically this does not correspond to a state of 
constant temperature, as would intuitively be expected from a 
non-changing state of a system. Instead of a constant temperature 
throughout the object, a constant \emph{flow of temperature} is 
achieved, an invariable gradient of temperature 
supported by the gravitational forces within the compact object, 
the inner parts being hotter than the outer parts. This corresponds 
to what is called the gravothermal catastrophe 
\cite{1968Lynd,1980Lynd,1985Anto}, which occurs in systems with 
negative heat capacities, such as is the case in gravitating 
fluids. The possibility of negative heat capacities was already 
considered very early by Maxwell \cite{1876Maxwell}. 
A negative heat capacity means that adding energy to the 
system leads to a cooling on it, and vice versa, a system 
becomes hotter by giving energy to a colder reservoir. Thus, the 
inner parts of a compact object heat up by losing energy to the 
cooler, outer parts, which can expand uncontrollably while the 
core becomes denser, and in principle there is nothing that can 
counteract this reaction until the gravitational collapse of the 
core, i.e., the gravothermal catastrophe. 
These considerations show that thermodynamical principles (and the 
statistical mechanical methods behind them) have to be reconsidered 
in the presence of gravitational (or generally, long-range) 
forces. 

The problem of a self-gravitating gas has been considered in the 
literature before, within the framework of traditional 
Boltzmann-Gibbs (BG) statistics. In Ref.~\cite{1990Padmanabhan} a very 
detailed discussion of the topic was presented by Padmanabhan. 
Ref.~\cite{2016Vela} gives a broad and well-founded introduction 
to the problems arising in the analysis of the thermodynamics of 
gravitating systems as well, approaching the issue from 
several angles and introducing and evaluating several possible 
advances. One of the established problems of the thermodynamics of 
gravitational systems is with the thermodynamic limit, i.e., 
\begin{equation} \label{eq:usuaTDlim}
	N\to \infty \,, V \to \infty \,, \mathrm{~and~} 
	N/V = const. 
\end{equation}
In this limit, thermodynamic functions like the temperature or 
the heat capacity do not exist due to divergences, and thus the 
thermodynamics of the system cannot be calculated. A possible way 
to resolve the issues with the thermodynamic limit is to modify it. 
In Ref. \cite{2002VS} de Vega and Sanchez extensively analyze the 
self-gravitating gas in BG statistics and show that thermodynamic 
functions exist in the so-called dilute limit, where 
\begin{equation} \label{eq:dillim}
	N\to \infty\,, V \to \infty \,, \mathrm{~and~} 
	N/V^{1/3} = const. 
\end{equation}
In this case, the linear dimension $L=V^{1/3}$ of the system scales 
with $N$ instead of its volume, and thus the way in which $N$ and 
$V$ are taken to infinity with respect to each other is changed in 
the thermodynamic limit. However, while the dilute limit 
\eqref{eq:dillim} is applied in \cite{2002VS} in the definition of 
appropriate dimensionless variables it is disregarded in the 
equation of state, where the limit \eqref{eq:usuaTDlim} is employed. 
Due to these conflicting applications of thermodynamic limits, and 
the general question as to what is the correct statistical mechanics 
picture to analyze a gravitating system, we would like to explore 
the option of alternative statistics. 

In this work, we are thus going to describe a self-gravitating 
system using a generalized statistical approach as proposed by 
Tsallis, and explore the consequences for thermodynamical 
quantities and the thermodynamic limit of the system.

Tsallis statistics has first been introduced in the 1980's (see e.g. 
\cite{1988Tsallis,1999Tsallis,2009Tsallis,2004Tsallis}) following 
the idea that for systems with long-range interactions such as 
gravity, but also complex systems with short-range interactions 
such as glasses or  systems with dissipation, the statistical 
analysis within the framework of BG statistics should be replaced 
with a more suitable generalized statistics. 
In general, BG statistics works very well for systems with 
short-range interactions, Markovian processes and 
ergodic systems, with equal probability of each microstate 
in the system. For certain systems, as for example such involving 
long-range forces, however, ergodicity is not guaranteed, and 
strictly speaking, the BG ensemble statistics and its 
way to count microstates considering equiprobability of all states 
does not apply in these cases. Long-range interactions can cause 
the energy and the entropy of a system to be non-additive, as is 
well-known in the case of black holes, where the entropy scales 
with the surface area of the black hole, and not its volume, thus 
making the system non-extensive. 
A generalized statistics takes into account these peculiarities, 
and defines alternative notions of thermodynamic quantities, 
which then can be interpreted in the traditional BG sense. Also 
the definition of heat capacities, which turn out negative in 
conventional BG statistics, and other relevant thermodynamic 
response functions, should be revised. The statistics proposed 
by Tsallis is also known as non-extensive generalized statistics. 

This paper is structured as follows. In \secref{sec:basic} 
we will introduce the physical system we are investigating, 
clarify important notions and quantities of a self-gravitating 
gas of free particles, and go on to review the basic ideas and 
properties of the Tsallis non-extensive statistics which will 
be used to analyze the thermodynamics of the system. We also 
comment on previous works presenting the case of an ideal gas 
within Tsallis statistics. 
In \secref{sec:CE} and \ref{sec:MCE} we will discuss in 
detail the statistical mechanics and thermodynamics of the 
system in the canonical and microcanonical ensembles, respectively, 
deriving the most important thermodynamic equation of state 
variables and response functions, and calculating them in the 
limit of a weak gravitational field, i.e., in the dilute regime. 
In \secref{sec:conclusions} we will comment on the connection 
between the two ensembles and the issue of ensemble equivalence, 
as well as conclude and summarize our work.

\section{Principal notions of the physical system and the statistical analysis} 
\label{sec:basic}

In this section, we will introduce the physical model of a 
self-gravitating gas as a system of $N$ particles interacting 
with each other gravitationally; and we will define and clarify 
the basic notions and properties of such a system. We will also 
review the most important ideas and definitions of the Tsallis 
non-extensive statistics, such as the Tsallis probability 
distribution function, the generalized entropy $S_q$, the 
definition of thermodynamic ensembles in this statistics and 
the physically relevant thermodynamic quantities and their 
interpretation.

\subsection{Self-gravitating gas}

The self-gravitating gas is a system of $N$ 
particles interacting with each other only via Newtonian gravity, 
as analyzed e.g. in Ref. \cite{2002VS}. We write the potential as 
the superposition of the interaction between pairs of particles, 
also know as pairwise approximation, thus treating the potential 
as a series of two-body problems instead of an $N$-body problem. 
At short distances, a repulsive particle interaction is assumed. 
The interaction potential between two single particles in the 
system is thus proposed as 
\begin{equation} \label{eq:cutoffA}
	-\frac{1}{\left| \mb{q_i}-\mb{q_j} \right|_A} =
	\begin{cases}
	-\frac{1}{\left| \mb{q_i}-\mb{q_j} \right|} \,, &
	\left | \mb{q_i}-\mb{q_j} \right | \geq A \,, \\
	+1/A \,, &
	\left| \mb{q_i}-\mb{q_j} \right| \leq A \,,
	 \end{cases}
\end{equation}
and $A\ll L$ is a repulsive short-distance cut-off, small compared 
to the size $L$ of the system. The presence of the repulsive 
short-range interaction introduces a finite size of the particles, 
and prevents the unphysical case of the collapse of the system into 
one point. It was shown in previous works \cite{2002VS} that the 
parameter $A$ can be taken to zero safely in the quantities we 
consider.

The Hamiltonian of such a system is the sum of the kinetic term 
$\mathcal{T}$ of all constituents and the interaction potential 
$\mathcal{U}$, 
\begin{equation} \label{eq:H}
\mathcal{H} = \mathcal{T} + \mathcal{U} 
= \sum_{i=1}^{N}\frac{p_i^2}{2m}-Gm^2 
u \left(\left|\mb{q_i}-\mb{q_j}\right|\right);
\end{equation}
where $G$ is the gravitational constant, and $m$ the mass of an 
individual particle. The potential 
$u (\left| \mb{q_i}-\mb{q_j} \right |)$ has been defined as 
\begin{equation} \label{eq:Uint}
u (\left| \mb{q_i}-\mb{q_j} \right |) = \sum_{1\leq i<j \leq N} 
\frac{1}{\left | \mb{q_i}-\mb{q_j} \right |_A}.
\end{equation}
For the sake of simplicity, we will assume the system to be 
contained in a cubic box of side length $L$, which will ultimately 
be taken in the limit $L \rightarrow \infty$, or 
$V \rightarrow \infty$, respectively. 

The long-range nature of the gravitational interactions leads to 
integrals which cannot be resolved analytically, and thus 
we will apply approximate methods for a weak gravitational field, 
considering a dilute gas cloud, which will enable us to obtain the 
results for thermodynamic quantities as a combination of the ideal 
gas contribution and small gravitational corrections.

\subsection{Tsallis statistics}

The generalized statistics proposed by Tsallis 
\cite{1988Tsallis,1999Tsallis,2009Tsallis,2004Tsallis} has been 
introduced in order to take into account systems with non-extensive 
thermodynamic behaviour. This generalization of the usual BG 
statistics revises, among other things, the notion of entropy by 
modifying the way of counting microstates. Let us first discuss 
the usual notion of BG entropy, and then show the generalization 
to Tsallis' non-extensive entropy. Entropy in BG statistics is 
defined using Shannon's formula as 
\begin{equation}
	S_{BG} = -k_B \sum_{i=1}^{\Omega} p_i \ln p_i \,,
\end{equation}
where $p_i$ is the probability of the $i$-th microstate, and the 
usual normalization condition applies to the sum of all 
probabilities, 
\begin{equation}
	\sum_{i=1}^{\Omega} p_i =1 \,.
\end{equation}
Assuming equiprobability, i.e., that each microstate is equally 
probable, $p_i = 1/\Omega$, the well-known BG entropy 
\begin{equation}
	S_{BG} = k_B \ln \Omega 
\end{equation}
is obtained. The BG entropy is, amongst other things, extensive, 
which means that when adding up the entropies of two independent 
systems $A$ and $B$, the result is the direct sum of both, 
\begin{equation}
	S_{BG}(A+B) = S_{BG}(A) + S_{BG}(B) \,.
\end{equation}
In Tsallis' generalized statistics, the entropy is defined via 
a generalization of the logarithmic function, the so-called 
q-logarithm, 
\begin{equation} \label{eq:qlog}
	\ln_q x = \frac{x^{1-q}-1}{1-q} \,.
\end{equation}
This function is actually a power-law of the variable $x$, 
containing a free parameter $q$, and reduces to the usual natural 
logarithm in the limit $q \to 1$. Using this q-logarithm, a 
generalized q-entropy can be defined as 
\begin{equation} \label{eq:Sqmicro}
	S_q = -k_B \sum_{i=1}^{\Omega} p_i^q \ln_q p_i 
	= \frac{1-\sum_{i=1}^{\Omega} p_i^q}{1-q} \,.
\end{equation}
This entropy, in contrast to the BG entropy, is 
non-extensive. When adding the entropies of two independent 
systems $A$ and $B$, the result is not simply the sum of the 
two entropies, but also includes a cross-term, 
\begin{equation} \label{eq:EntSum}
	S_q(A+B) = S_q(A) + S_q(B) + (1-q) S_q(A) S_q(B) \,.
\end{equation}
This cross-term is associated with the interaction between the 
two systems, is labeled by the so-called non-extensivity 
parameter $q$ and makes the q-entropy non-extensive. 
Depending on the value of the parameter $q$, the q-entropy is 
superextensive for $q<1$, or subextensive for $q>1$, while for 
$q=1$ the extensivity of the BG form is recovered. Introduced 
as a free parameter, $q$ should be assigned a specific value 
depending on the system in question, and its properties with 
respect to the type of interactions between the constituents. 
In the presence of gravitational interaction, $q<1$ should be 
expected, i.e., a superextensive entropy. This can be illustrated 
in the simple case of a Schwarzschild black hole, where the 
entropy of a system with the mass $M_1 + M_2$ is higher than 
the sum of the entropies of two black holes with mass $M_1$ and 
$M_2$, and has as well been argued in the case of two particular 
astrophysical scenarios in Ref.\,\cite{2016Vela}.

As the generalized logarithm, also a generalized q-exponential 
function can be defined \cite{2009Tsallis} as its inverse, 
\begin{equation} \label{eq:qexp}
	exp_q (x) = \Big[ 1+ (1-q) x \Big]^{\frac{1}{1-q}} \,,
\end{equation}
which equally recovers the usual exponential function for 
$q \to{1}$. Calculations become more involved for these generalized 
functions, as some of the usual identities for exponential and 
logarithm change, such as the q-exponential of a sum of arguments 
or the q-logarithm of a product of arguments. For a collection of 
identities regarding the $q$-calculus, see the appendix of Ref. 
\cite{2009Tsallis}. 

Besides the generalized entropy, also other notions have to be 
generalized. Depending on the ensemble, different quantities such 
as the internal energy, the free energy, temperature or pressure 
are obtained in an analogous way to the BG statistics, but based 
on the Tsallis statistics and its respective distribution 
functions. 

Notions change only slightly for the microcanonical ensemble. The 
entropy of the system is calculated from \eqeqref{eq:Sqmicro}. 
After the maximization of the entropy under usual constraints such 
as constant internal energy $E$, volume $V$ and number of particles 
$N$, the equilibrium distribution turns out to be a constant, i.e., 
an equiprobability distribution, and so in analogy to the BG case 
the entropy is \cite{2009Tsallis}
\begin{equation} \label{eq:STsallis}
	S_q = k_B \ln_q \Omega \,,
\end{equation}
with $\Omega$ being the number of possible microstates of the 
system, just as in BG statistics. From this entropy, the 
thermodynamic equations of state are calculated -- and lead to the 
Tsallis temperature as 
\begin{equation} \label{eq:Tq}
	T_q \equiv \frac{1}{k_B \beta} = \left( \frac{\partial S_q}
	{\partial E} \right)_{V,N}^{-1} \,,
\end{equation}
and the Tsallis pressure as 
\begin{equation} \label{eq:Pq}
	P_q = T_q \, \left( \frac{\partial S_q}{\partial V} 
	\right)_{E,N} \,.
\end{equation}
Here, $\beta$ is the inverse temperature directly derived from 
the Tsallis temperature $T_q$. However, based on considerations 
of a generalized notion of equilibrium (thermal and mechanical), 
derived from the generalized rule of addition 
\eqref{eq:EntSum} for the Tsallis entropy, the Tsallis intensities 
have been debated as the true physical intensive quantities of the 
system. In \cite{2001Abe,2003Toral}, it is argued that instead 
of the intensive Tsallis variables \eqref{eq:Tq} and \eqref{eq:Pq} 
above, more physical definitions of intensive variables should be 
interpreted as the true intensive quantities of the system, i.e., 
\begin{equation} \label{eq:Tphys}
	T^* = \left( 1+ \frac{1-q}{k_B} S_q \right)
	\left( \frac{\partial S_q}{\partial E} \right)_{V,N}^{-1} 
	=: \frac{1}{k_B \beta_q} \,,
\end{equation}
and 
\begin{equation} \label{eq:Pphys}
	P^* = \frac{T^*}{1+ \left[ (1-q)/k_B \right] S_q}
	\left( \frac{\partial S_q}{\partial V} \right)_{E,N} \,.
\end{equation}
Here, the inverse physical temperature is defined as $\beta_q$, 
and the factor of proportionality between $\beta$ and $\beta_q$ 
is commonly referred to as $c$, i.e., 
\begin{equation} 
	c = 1+ \frac{1-q}{k_B} S_q \,,
\end{equation}
and 
\begin{equation} \label{eq:betaq}
	\beta_q = \frac{\beta}{c} \,.
\end{equation}
We will thus calculate these physical (starred) thermodynamic 
variables instead of the original Tsallis ones in our analyses. 

In contrast to the microcanonical ensemble, where everything is 
derived from the entropy and only thereafter a partition function 
is introduced, in the canonical ensemble everything 
originates from the canonical partition function $Z_q$. These 
partition functions for microcanonical or canonical ensembles 
are defined in analogy to their respective BG partition functions, 
with slight modifications, and will be introduced in the analyses 
in sections \ref{sec:CE} and \ref{sec:MCE}. In the canonical 
ensemble, also extensive variables are referred to as Tsallis 
variables, e.g. the free energy $F_q$. The entropy is connected 
to the canonical partition function via 
\begin{equation} \label{eq:SqCE}
	S_q = k_B \ln_q Z_q \,,
\end{equation}
and also the Tsallis extensive variables have definitions involving 
their physical intensive counterparts, such as the physical free 
energy \cite{2001Abe,2003Toral}, 
\begin{equation} \label{eq:FphysCE}
	F^* = -k_B T^* \ln Z_q \,.
\end{equation}
From this definition, the physical pressure can be derived in the 
case of the canonical ensemble as 
\begin{equation} \label{eq:PphysCE}
	P^* = -\left( \frac{\partial F^*}{\partial V} \right)_{T^*} \,,
\end{equation}
i.e., the derivative of the physical free energy with respect to 
the volume at constant physical temperature. This is the alternative 
derivation of the pressure for the canonical ensemble. 
In \secref{sec:conclusions} we will comment on this question in the 
context of our own calculations for microcanonical and canonical 
ensemble. 

Additionally, the thermodynamic limit is modified in Tsallis' 
statistics. In usual BG statistics, extensive quantities are scaled 
with respect to the total number of particles according to 
\eqref{eq:usuaTDlim}, and intensive quantities do not scale at 
all, but remain constant in all of the system for the 
thermodynamic limit $N,V\to \infty$. In Tsallis statistics, the 
notions of extensive and intensive variables are generalized to 
three classes: extensive variables should be differentiated into 
real extensive quantities, such as entropy and volume, which 
indeed scale with the number of particles in the system, and 
pseudo-extensive ones, like the internal energy or the free 
energy, which should be 
scaled with a modified factor of $N \, N^*$ in the thermodynamic 
limit. Here, $N^* \propto N^{1-\alpha/d}$, $\alpha$ being the 
exponent of the interaction potential (in the case of gravity 
$\alpha =1$), and $d$ the spatial dimension of the system (in this 
case $d=3$), and the proportionality constant is a combination of 
$\alpha$ and $d$. The scaling of pseudo-extensives thus takes into 
account the particular type of long-range interaction and other 
properties of the system. In the case of the self-gravitating 
gas, $N^*\propto N^{2/3}$. Intensive quantities should then be 
considered pseudo-intensive, and scale with the factor $N^*$. 
The thermodynamic limit in our case for a large system with 
$N \to \infty$ thus will be carried out for the extensives as 
\begin{equation} \label{eq:LimExt}
	\lim_{N,V\rightarrow \infty} \frac{S}{N} \,,\frac{V}{N} 
	= const \,,
\end{equation}
for the pseudo-extensives as 
\begin{equation} \label{eq:LimPsExt}
	\lim_{N\rightarrow \infty} 
	\frac{F^*}{N \, N^*} = const \,,
\end{equation}
and for the pseudo-intensities as 
\begin{equation} \label{eq:LimPsInt}	
	\lim_{N \rightarrow \infty} \frac{T^*}{N^*} \,,
	\frac{P^*}{N^*} = const \,.
\end{equation} 

Previous works exist on the case of an ideal gas of particles 
analyzed via Tsallis statistics \cite{1999Abe,2006Parv}. In the case 
of the canonical ensemble of the ideal gas, we obtain coincidence 
between their results \cite{1999Abe} and ours in the limit 
of zero gravity. In the case of the microcanonical ensemble 
however, we differ from \cite{2006Parv} in methodology. In order 
to achieve convergent results, \cite{2006Parv} promoted the 
parameter $q$ to a thermodynamic variable, scaling as $1/N$ in 
the thermodynamic limit. We do not adopt this idea, but retain 
$q$ as a non-determined constant parameter in the theory, and 
thus do not reproduce the results of \cite{2006Parv} for the 
microcanonical ensemble in the limit of zero gravity.

\section{Tsallis non-extensive statistical mechanics of the self-gravitating gas: the canonical ensemble}
\label{sec:CE}
We first consider the canonical ensemble for the Tsallis analysis 
of the self-gravitating gas, and start with the Tsallis canonical 
partition function. The generalization to q-statistics consists in 
the use of the q-exponential of the Hamiltonian \eqref{eq:H} in the 
integral over coordinate and momentum space, 
\begin{equation}
	Z_q = \frac{1}{N!h^{3N}}\int d^{3N}q \, d^{3N}p \, 
	\exp_q{\left(-\beta_q\mathcal{H}(\textbf{p},\textbf{q})
	\right)} \,,
\end{equation}
where $\beta_q$, the inverse physical temperature, is used as the 
Lagrange multiplier associated with the energy constraint in 
Tsallis' nonextensive statistical mechanics \cite{2001Abe}. This 
quantity is functionally different from $\beta$, the Lagrange 
multiplier associated with the energy constraint in BG statistical 
mechanics, and was introduced in \eqref{eq:betaq}. 

The integral can be evaluated via the introduction of the complex 
plane representation of the Gamma function \cite{1995Prato}, 
\begin{equation}
	\frac{1}{\Gamma(y)} = \frac{i}{2\pi} x^{1-y} \oint_C (-t)^{-y} 
	e^{-tx} dt \,,
\end{equation}
for non-integer $y$, with the identifications 
\begin{equation}
	x = 1 - (1-q) \beta_q \mathcal H \quad \textrm{and} 
	\quad y-1 = \frac{1}{1-q} \,.
\end{equation}
This substitution holds for $q<1$, i.e., superextensive 
entropies such as is the case for a self-gravitating gas.
The integral over $d^{3N}p$ can then be calculated using 
$DN$-dimensional polar coordinates, and leads to 
\begin{eqnarray}
	Z_q &=& \frac{\Gamma \left( \frac{2-q}{1-q} \right)}{N! 
	\, h^{3N}} \left( \frac{2\pi m}{(1-q)\beta_q}\right)^{3N/2} 
	\frac{i}{2\pi} \\
	&& \oint_C dt (-t)^{-\frac{2-q}{1-q}-\frac{3N}{2}} 
	e^{-t} \int d^{3N}q \, e^{t(1-q)\beta_q \mathcal{U}} \,,
	\nonumber
\end{eqnarray}
where $\mathcal{U}$ has been defined in \eqeqref{eq:H}. 
In order to simplify the integral over the coordinates $\mb{q}_i$, 
they are rescaled such that the limits of the integration do not 
depend on the size of the box, i.e., $\mb{q}_i = L \mb{r}_i$. Thus, 
the integral measure becomes 
\begin{equation} \label{eq:rescaling}
	d^3 q_i = L^3 d^3 r_i \,.
\end{equation}
Further, we will introduce a new dimensionless variable $\eta$,       
\begin{equation} \label{eq:eta}
	\eta = \frac{G m^2 N\beta_q}{L} \,,
\end{equation}
which is constructed to be a truly intensive quantity remaining 
constant in the Tsallis thermodynamic limit as defined in 
\eqref{eq:LimExt}--\eqref{eq:LimPsInt}. It is the same variable 
$\eta$ as used in \cite{2002VS}, and under their modified 
thermodynamic limit, $\eta$ remains a constant as well. \\
Using these substitutions, the partition function can be written as  
\begin{eqnarray} \label{eq:Zq1}
	Z_q &=& \frac{V^N \Gamma \left( \frac{2-q}{1-q} \right) }{N!\, 
	h^{3N}} \left( \frac{2\pi m}{(1-q)\beta_q}\right)^{3N/2} 
	\frac{i}{2\pi} \\
	&& \oint_C dt (-t)^{-\frac{2-q}{1-q}-\frac{3N}{2}} e^{-t} 
	\int d^{3N}r \, e^{\eta_q u \left( | \mb{r_i}-\mb{r_j} 
	| \right)} \,, \nonumber
\end{eqnarray}
where $u\left( | \mb{r_i}-\mb{r_j} | \right)$ is the interaction 
potential \eqref{eq:Uint} as defined before, but in terms of the 
new variable $\mb{r}$, and for brevity we have introduced 
\begin{equation}
	\eta_q = \frac{(-t) (1-q)}{N} \eta \,.
\end{equation} 
In order to be able to further process this 
expression, in particular to carry out the coordinate integrals, 
we now resort to an approximation approach, considering the regime 
of weak gravity, i.e., a dilute self-gravitating gas, to expand the 
interaction potential $u\left( | \mb{r_i}-\mb{r_j} | \right)$ and 
the exponential term.

\subsection{Dilute regime}
The dilute regime assumes a weakly gravitating gas with low density. 
The introduced quantity $\eta$ remains constant in the Tsallis 
thermodynamic limits \eqref{eq:LimExt}--\eqref{eq:LimPsInt}, but 
can be assumed to be small in the dilute regime, i.e., $\eta \ll 1$. 
In that case, a Taylor expansion of the interaction potential 
can be carried out, i.e., up to second order, 
\begin{equation} \label{eq:expansionEq}
	e^{\eta_q u(\left| \mb{r_i}-\mb{r_j} \right|)} 
	\approx 1 + \eta_q u + \frac{1}{2!} \eta_q^2 u^2 \,.
\end{equation}
Furthermore, restricting the potential interactions between $N$ 
particles to a sum of identical two-body interactions, the potential 
can be written as    
\begin{eqnarray} \label{eq:u}
	u(\left| \mb{r_i}-\mb{r_j} \right|) 
	&=& \frac{1}{\left| \mb{r_1}-\mb{r_2} \right|} 
	\left[ (N-1)+(N-2)+\ldots +1 \right] \nonumber \\
	&=& \frac{1}{\left| \mb{r_1}-\mb{r_2} \right|} 
	\sum_{k=1}^{N}(N-k) \,.
\end{eqnarray}
In the limit $N \rightarrow \infty$ this results in 
\begin{equation} \label{eq:usimple}
	u(\left| \mb{r_i}-\mb{r_j} \right|) = \frac{N(N-1)}{2 
	\left| \mb{r_1}-\mb{r_2} \right|} \,.
\end{equation}
The quadratic term in \eqeqref{eq:expansionEq} can be simplified in 
a similar manner, assuming that the contributing terms stem from 
the interaction between two independent particles squared, from 
interactions between one particle separately with two others, or 
from interactions between two pairs of particles with four 
independent particles. Thus, the second order interaction in 
\eqeqref{eq:expansionEq} can be written as \cite{2002VS}
\begin{eqnarray} \label{eq:uquad}
	u(\left| \mb{r_i}-\mb{r_j} \right|))^2 = &&
	\frac{N(N-1)}{2 \left| \mb{r_1}-\mb{r_2} \right|^2} + 
	\frac{N(N-1)(N-2)}{\left| \mb{r_1}-\mb{r_2} \right| 
	\left| \mb{r_1}-\mb{r_3} \right|} \nonumber\\
	&& + \frac{N(N-1)(N-2)(N-3)}{4\left| 
	\mb{r_1}-\mb{r_2} \right| \left| \mb{r_3}-\mb{r_4} \right|} \,. 
\end{eqnarray}
Plugging all of these things into the coordinate integral of 
\eqeqref{eq:Zq1}, up to second order we end up with 
\begin{equation} \label{eq:Zapprox}
	e^{\eta_q u(\left| \mb{r_i}-\mb{r_j} \right|))}\approx 
	1 - tN^2A + t^2N^2B \,,
\end{equation}
where we introduced 
\begin{eqnarray} \label{eq:AB}
A&=&\frac{\eta (1-q)b_0}{2N}\left(1-\frac{1}{N} \right) \,, \\
B&=&\frac{\eta^2 (1-q)^2}{2N^2} \Bigg[ 
\frac{b_0^2}{4}\left(1-\frac{1}{N} \right) 
\left(1-\frac{2}{N} \right)\left(1-\frac{3}{N} \right) \nonumber\\
&& + \frac{b_2}{2N^2}
\left(1-\frac{1}{N} \right) + \frac{b_1}{N} \left(1-\frac{1}{N} 
\right)	\left(1-\frac{2}{N} \right) \Bigg] \,, ~~~~~
\end{eqnarray}
and further defined the gravitational integrals as 

\begin{eqnarray}
	b_0&&=\int _0^1 d^{3}r_1 d^{3}r_2 \,
		\frac{1}{\left| \mb{r_1}-\mb{r_2} \right|} \,, \\
	b^2_0&&=\int _0^1 d^{3}r_1 
	d^{3}r_2 d^{3}r_3 d^{3}r_4 \, 
		\frac{1}{\left| \mb{r_1}-\mb{r_2} \right|} 
		\frac{1}{\left| \mb{r_3}-\mb{r_4} \right|} \,, \\
	b_1&&= \int _0^1 d^{3}r_1 d^{3}r_2 d^{3}r_3 
		\, \frac{1}{\left| \mb{r_1}-\mb{r_2} \right|} 
		\frac{1}{\left| \mb{r_1}-\mb{r_3} \right|} \,, \\
	b_2&&= \int _0^1 d^{3N}r_1 d^{3N}r_2 \, 
		\frac{1}{\left| \mb{r_1}-\mb{r_2} \right|^2} \,. 
\end{eqnarray}
These coefficients are simple expressions which can be computed 
numerically and depend on the symmetry of the box containing the 
system. They do not carry any dependence on variables anymore. 
The short-distance cutoff as introduced in \eqref{eq:cutoffA} 
needs to be considered only in these coefficients, and it has 
been shown in \cite{2002VS} that its influence on the numerical 
results can be neglected. With these substitutions, we can 
compute the integral over the auxiliary variable $t$ as in 
\cite{1995Prato}, leading to the partition function in the 
dilute regime up to second order as 
\begin{widetext}
\begin{eqnarray} \label{eq:LimZq}
	Z_q &\simeq& \left[\frac{2\pi m}{(1-q)\beta_q}\right]^{3N/2} 
	\frac{V^N}{N!\, h^{3N}} \frac{\Gamma\left(\frac{2-q}{1-q} 
	\right)}{\Gamma\left(\frac{2-q}{1-q}+\frac{3N}{2}\right)} 
	\cdot \left[ 1 + AN^2 \frac{\Gamma\left( \frac{2-q}{1-q}+
	\frac{3N}{2} \right)}{\Gamma \left( \frac{1}{1-q}+\frac{3N}{2} 
	\right)} + BN^2 \frac{\Gamma\left( \frac{2-q}{1-q}+\frac{3N}{2} 
	\right)} {\Gamma \left( \frac{q}{1-q}+\frac{3N}{2}\right)} 
	\right] \nonumber \\
	&=& Z_q^{(IG)} \cdot Z_q^{(grav)} \,. 
\end{eqnarray}
\end{widetext}
Already here we can see that in the limit of $\eta =0$, i.e., 
in the absence of gravitational interaction, $A=B=0$, the 
partition function recovers the result of the ideal gas in 
Tsallis statistics according to \cite{1999Abe}, 
\begin{equation} \label{eq:ZqIG}
	Z_q^{(IG)} = \left[\frac{2\pi m}{(1-q)\beta_q}\right]^{3N/2} 
	\frac{V^N} {N!\, h^{3N}} \frac{\Gamma\left(\frac{2-q}{1-q}
	\right)}{\Gamma\left(\frac{2-q}{1-q}+\frac{3N}{2}\right)} \,.
\end{equation}
Note that the result does not recover the partition function of 
the ideal gas in BG statistics -- there is an additional fraction 
of two $\Gamma$-functions. The limit $q\to 1$ of the above 
expression \eqref{eq:ZqIG} does not straightforwardly reproduce the 
desired result. We thus argue that the limit $q\to 1$ can not be 
taken at any point in these calculations, particularly since the 
introduction of the integral transformations according to 
\cite{1995Prato} is not applicable in the case $q = 1$. Our 
result for $Z_q^{(IG)}$ does however coincide with the result 
in \cite{1999Abe} for the ideal gas in the $q$-calculus. 

We are now interested in the thermodynamic limit $N,V\to \infty$ 
under consideration of the Tsallis rules for the limit 
\eqref{eq:LimExt}--\eqref{eq:LimPsInt}. As mentioned before, 
the quantity $\eta$ was constructed to remain constant in 
this limit. Considering the arguments of the Gamma functions, we 
identify 
\begin{equation}
	\frac{2-q}{1-q} +\frac{3N}{2} = \frac{1}{1-q} +\frac{3N}{2}+1 
	= \frac{q}{1-q} +\frac{3N}{2} +2 \,.
\end{equation}
For $N\rightarrow \infty$, it is justified to use the 
approximation 
\begin{equation}
	\frac{3N}{2} \approx \frac{3N}{2}+ 1 \approx \frac{3N}{2}+2 \,,
\end{equation}
and thus we can simplify, neglecting higher order terms, to 
\begin{equation}
	\ln Z_q^{(grav)} \simeq \ln{ \left[ 1 + AN^2 + BN^2 
	\right]} \,.
\end{equation}
For the dilute regime, we can approximate the logarithm 
to second order as 
\begin{equation}
	\ln(1+x)\approx x-\frac{x^2}{2!} \,.
\end{equation}
With this, we can express the thermodynamic limit of the expression 
$\ln{Z_q^{(grav)}}/N$ as 
\begin{eqnarray} \label{eq:ZgravLimit}
	&& \lim_{N\to\infty} \frac{1}{N}\ln{Z_q^{(grav)}} \simeq \\
	&& ~~~~~~~ \frac{\eta (1-q) b_0}{2} + \eta^2 (1-q)^2 
	\left(\frac{b_1}{2} - \frac{b_0^2}{2} \right)
	\,. \nonumber
\end{eqnarray}
In the following, we calculate thermodynamically relevant 
quantities, such as the equation of state and heat capacity 
in the dilute regim.

\subsection{Thermodynamic equation of state and heat capacity}
In the canonical ensemble, the physical temperature $T^*$, or 
equivalently, the inverse physical temperature $\beta_q$, is fixed 
due to the energy constraint, and enters the calculation as a 
constant Lagrange multiplier. In contrast, the physical pressure 
\eqref{eq:PphysCE} can be calculated from the physical free energy 
\eqref{eq:FphysCE}, by derivation of the canonical partition 
function as obtained in the dilute regime, \eqeqref{eq:LimZq}. 
The physical pressure divided by the physical temperature is 
thus 
\begin{equation} \label{eq:PphysCanEn}
	\frac{P^*}{k_B T^*} = \left( 
	\frac{\partial \ln Z_q}{\partial V} \right)_{T^*} \,,
\end{equation}
i.e., the derivative of the logarithm of the partition function, 
taken at constant $\beta_q$. The derivative of the ideal gas part 
of the partition function yields 
\begin{equation}
	\frac{\partial \ln Z_q^{(IG)}}{\partial V} = \frac{N}{V} \,,
\end{equation}
thus recovering the analog of the ideal gas equation of state in 
the limit of zero gravitational interactions, i.e., the equation 
of state of the ideal gas in the Tsallis statistics, using the 
Tsallis variables $T^*$ and $P^*$. The equation of state of the 
gravitating gas can be written as 
\begin{equation} \label{eq:PphysDilReg}
	\frac{P^* V}{N k_B T^*} = 
	1 - \frac{\eta}{3N} \frac{\partial}{\partial \eta} 
	\ln Z_q^{(grav)} \,.
\end{equation}
Substituting \eqeqref{eq:ZgravLimit} into \eqeqref{eq:PphysDilReg}, 
interchanging the thermodynamic limit and the derivative with 
respect to $\eta$, allows us to compute the thermodynamic limit 
of the equation of state. This limit can be taken without any 
additional factors of $N$ on the left hand side, since the weights 
of $P^*$ and $T^*$ cancel each other, and also the ratio $V/N$ 
represents already the correct thermodynamic limit. The equation 
of state up to second order thus results in 
\begin{equation}\label{eq:LimPq}
	\frac{P^* V}{N k_B T^*} \simeq 1 - 
	\frac{\eta (1-q)b_0}{6} -\frac{\eta^2 (1-q)^2}{3} 
	\left(b_1 - b_0^2 \right) 
	\,.
\end{equation}
This result is the same as obtained in 
\cite{2002VS}, considering the differences in the definition 
of the $b_i$ and the use of the Tsallis variables instead of 
the usual BG ones. Moreover, there is an explicit dependence 
on the parameter $q$.

While the equation of state is useful in order to study the 
particular features of any thermodynamic system, relevant 
information often appears when studying the second derivatives 
of a certain thermodynamic potential, the so-called response 
functions. Thus it is also instructive to consider the specific 
heat capacity at constant volume, defined in the Tsallis statistics 
as \cite{2001Abe}
\begin{equation} \label{eq:cVcan}
	(c_{V})_q= - \frac{T^*}{N} \left( \frac{\partial^2 F^*}
	{\partial T^*} \right)_V \,.
\end{equation}
Using the definition \eqref{eq:FphysCE} of the physical free energy, 
and in turn the result for the partition function \eqref{eq:LimZq} 
combined with the thermodynamic limit of its gravitational part from 
\eqref{eq:ZgravLimit}, the specific heat capacity to second order 
turns out as 
\begin{equation} \label{eq:Limcv}
	\frac{(c_{V})_q}{k_B} \simeq \frac{3}{2} 
	+ \eta^2 (1-q)^2 \left(b_1- b_0^2 \right)
	\,.
\end{equation}
In the case of zero gravity, $\eta=0$, the result of the ideal 
gas is reproduced in the analogous Tsallis framework, since the 
definition of the canonical ensemble includes the use of Tsallis 
temperature. 
The specific heat capacity for the ideal gas is 
positive, and the gravitational corrections show that it remains 
positive also in the presence of gravitational forces. This is 
qualitatively the same result as obtained in \cite{2002VS}, 
where the heat capacity was not calculated explicitly, but it was 
shown that it is always positive for the canonical ensemble.

\section{Tsallis non-extensive statistical mechanics of the self-gravitating gas: the microcanonical ensemble}
\label{sec:MCE}
To have a more complete perspective of the self-gravitating gas 
in Tsallis non-extensive statistical mechanics, it is desirable 
to explore other ensembles in order to compare the obtained 
thermodynamic properties. As in regular BG statistics, the 
microcanonical ensemble is the most fundamental one, where the 
system is considerd to be closed in all regards. The microcanonical 
entropy of the system is given as in \eqeqref{eq:STsallis}, as the 
q-logarithm of the number of accessible microstates of the system, 
\begin{equation}\label{eq:STsallisMCE}
	S_q = k_B \ln_q \Omega (E,V,N) \,,
\end{equation}
for a given energy $E$, volume $V$ and particle number $N$ of the 
gas \cite{2009Tsallis}. These microstates are restricted by the 
energy constraints imposed by the Hamiltonian \eqref{eq:H}. 

We follow the standard procedure to obtain the microstates of the system. 
In general, the number of total microstates for a specific energy 
constraint is given as
\begin{equation}
	\omega (E)=\int_{\mathcal{H}\leq E} d^{3N}q\ d^{3N}p \,.
\end{equation}
Thus, the density of microstates at an energy $E$ is given by 
\begin{equation} 
	\Omega (E) = \frac{1}{N! h^{3N}} 
	\frac{\partial \omega(E)}{\partial E} \,.
\end{equation}
Using the Hamiltonian \eqref{eq:H}, the momentum integral can be 
calculated as 
\begin{equation} \label{eq:intp}
	\int_{\mathcal{H}\leq E} d^{3N}p=\frac{\pi^{3N/2}}{\Gamma 
	\left(\frac{3N}{2}+1\right)} 
	\Big[2m(E - \mathcal{U} )\Big]^{3N/2} \,,
\end{equation}
which is the volume of a $3N$--dimensional hypersphere of radius 
$2m(E-\mathcal{U})$, and $\mathcal{U}$ has been defined in 
\eqref{eq:H}. The number of microstates for an energy $E$ is then 
given by derivation of \eqref{eq:intp} as 
\begin{equation}\label{eq:Omega}
	\Omega = \frac{(2\pi m)^{3N/2}}{N! \,h^{3N}
	\Gamma\left(\frac{3N}{2}+1\right)}
	\int d^{3N}q\ \big[E-\mathcal{U}\big]^{3N/2-1} \,. 
\end{equation}
For an ideal gas, i.e., without any particle interactions, the 
number of microstates can be computed from \eqeqref{eq:Omega} 
assuming $\mathcal U =0$ as 
\begin{equation} \label{eq:OmIG}
	\Omega^{(IG)}=\frac{(2\pi m)^{3N/2}V^N}{N! h^{3N}
	\Gamma \left(\frac{3N}{2}+1\right)} E^{3N/2-1} \,.
\end{equation}
Therefore, it is useful to separate the number of microstates 
\eqref{eq:Omega} in two different contributions,  
\begin{equation} \label{eq:OmIG+Z}
	\Omega=\Omega ^{(IG)} \cdot \mathcal{Z}^{(grav)} \,,
\end{equation}
one from the ideal gas behaviour of the kinetic term 
$\Omega^{(IG)}$, and one due to the gravitational interactions. 
$\mathcal{Z}^{(grav)}$ is the configuration integral containing 
the gravitational interaction, a sort of microcanonical partition 
function, given by 
\begin{equation}\label{eq:ZCI}
	\mathcal{Z}^{(grav)}=\frac{1}{V^NE^{3N/2-1}}\int d^{3N}q 
	\left[E-\mathcal{U}(\left| \mb{q_i}-\mb{q_j} \right|)
	\right]^{3N/2-1} \,.
\end{equation} 
In order to carry out the integral over the coordinates $\mb{q}_i$, 
we introduce the same rescaled coordinate $\mb{r}_i$ as in the case 
of the canonical ensemble, c.f. \eqeqref{eq:rescaling}. Further, we 
will introduce a new dimensionless variable $\chi$, 
\begin{equation} \label{eq:chi}
	\chi = \frac{EL}{Gm^2 N^2} \,. 
\end{equation}
which is an intensive quantity constructed in such a way that it 
remains constant in the Tsallis thermodynamic limit as defined by 
Eqs. \eqref{eq:LimExt}--\eqref{eq:LimPsInt}. Again, it is the same 
variable as $\chi$ in Ref. \cite{2002VS}, where it was a constant 
under their modified thermodynamic limit. With these substitutions, 
the configuration integral can be rewritten as 
\begin{eqnarray} \label{eq:ZCIchi}
	&& \mathcal{Z}^{(grav)} = \chi^{1-3N/2} \\
	&& ~~~~~~~~~~\cdot \int _0^1 \cdots \int _0^1 
	d^{3N} \mb{r}_i \left[ \chi + \frac{1}{N^2} u(\left| 
	\mb{q_i}-\mb{q_j} \right|) \right]^{3N/2-1} \,,\nonumber
\end{eqnarray} 
leading to the Tsallis entropy 
\begin{equation} \label{eq:STsallisSep}
	S_q=k_B\ln _q \left[ \Omega ^{(IG)}\cdot \mathcal{Z}^{(grav)} 
	\right]\,. 
\end{equation}
We will now introduce again  the case of weak gravity, before we 
then calculate thermodynamic state functions such as the equation 
of state and the heat capacity.

\subsection{Dilute regime}
As in the canonical ensemble, we will now define the regime of 
weak gravity, i.e., the case of low densities, and then consider 
the thermodynamic limit. Let us rewrite $\mathcal{Z}^{(grav)}$ 
\eqref{eq:ZCIchi} in the form 
\begin{eqnarray}
	&& \mathcal{Z}^{(grav)} = \\
	&& ~~ \int _0^1 \cdots \int _0^1 d^{3N} 
	\mb{r}_i \left[ 1 + \frac{1}{\chi N^2} \sum_{1\leq i<j \leq N}
	\frac{1}{\left| \mb{r_i}-\mb{r_j} \right|}
	\right]^{3N/2-1} \,. \nonumber
\end{eqnarray} 
The dilute limit, where $\chi \gg 1$, implies $1/\chi \ll 1$, and 
thus we use a Taylor expansion up to second order as 
\begin{eqnarray}
	(1+x)^{\frac{3N}{2}-1} &\approx& 1 + 
	\left( \frac{3N}{2}-1 \right) x \\ 
	&& + \left( \frac{3N}{2}-1 \right) 
	\left( \frac{3N}{2}-2 \right) \frac{x^2}{2!} 
	\,. \nonumber
\end{eqnarray} 
for the argument 
\begin{equation}
	x = \frac{1}{\chi N^2} u(\left| \mb{r_i}-\mb{r_j} \right|) \,.
\end{equation}
Thus, we have up to second order, 
\begin{eqnarray} \label{eq:expansionZ}
	&& \left[ 1+ \frac{u(\cdot)}{\chi N^2}\right]^{3N/2-1} \approx 
	1 + \left( \frac{3N}{2}-1 \right) \frac{u(\cdot)}{\chi N^2} 
	~~~~~\\
	&& ~~~~~~~~~~~~~~~~~~~ + 
	\frac{1}{2}\left( \frac{3N}{2}-1 \right) 
	\left( \frac{3N}{2}-2 \right) \frac{u(\cdot)^2}{\chi ^2N^4} 
	\,.\nonumber
\end{eqnarray}
The potential $u(\cdot) = u(\left| \mb{r_i}-\mb{r_j} \right|)$, as 
defined in \eqref{eq:Uint}, can be reduced in this regime using the 
same symmetry arguments presented for the canonical ensemble from 
\eqeqref{eq:u} through \eqeqref{eq:uquad}, in order to obtain the 
corresponding contribution to each term of the expansion up to the 
second order in inverse $\chi$. Therefore, the same virial 
coefficients $b_0, b^2_0, b_1$ and $b_2$ appear for the 
microcanonical ensemble, as defined earlier. The logarithm of 
$\mathcal{Z}^{(grav)}$ in the limit $N\to\infty$ up to second order 
in $1/\chi$ is then written as 
\begin{equation} \label{eq:Zcl-limit}
	\lim_{N\to\infty}\frac{1}{N}\ln\mathcal{Z}^{(grav)} \simeq 
	- \frac{9b_0}{2\chi} + \frac{9}{8 \chi^2} (b_1-42b_0^2) 
	\,. 
\end{equation} 
This result can be used in the following in order to simplify 
the computation of the thermodynamic state functions.

\subsection{Thermodynamic equation of state and heat capacity}
We proceed to compute thermodynamic quantities from the 
entropy of the system according to the definitions of 
physical thermodynamic quantities given in the introduction 
\cite{1999Abe,2003Toral} -- the temperature \eqref{eq:Tphys} 
and the pressure \eqref{eq:Pphys}, to be combined to yield 
the equation of state, as done in the canonical ensemble. We 
will first calculate the equations of state in an abstract form, 
before investigating it in the dilute regime to obtain concrete 
expressions to compare to the results of the canonical ensemble. 

In order to calculate the physical temperature, it is necessary 
to compute the derivative of the microcanonical Tsallis entropy 
$S_q$ \eqref{eq:STsallisSep}, given by 
\begin{eqnarray} \label{eq:SqDerE}
	\left( \frac{\partial S_q}{\partial E} \right)_{V,N} 
	&=& k_B  \left[ \Omega ^{(IG)}\cdot \mathcal{Z}^{(grav)} 
	\right]^{1-q} \\
	&& ~~~ \cdot \left[\frac{\partial}{\partial E} \ln{\Omega^{(IG)}} 
	+ \frac{\chi}{E} \frac{\partial}{\partial\chi} 
	\ln{\mathcal{Z}^{(grav)}} \right] \,. \nonumber
\end{eqnarray}
Due to the properties of the q-logarithm, the prefactor in these 
expressions can be identified as 
\begin{equation}
\left[ \Omega ^{(IG)}\cdot \mathcal{Z}^{(grav)} \right]^{1-q} = 
1+ \frac{1-q}{k_B} S_q = c \,,
\end{equation}
i.e., the prefactor from the derivation cancels with the prefactor 
that should be included into the physical thermodynamic variables 
\eqref{eq:Tphys} and \eqref{eq:Pphys}. 
Substituting relation \eqref{eq:SqDerE} into \eqeqref{eq:Tphys} 
and simplifying leads to 
\begin{equation} \label{eq:Tphysmicro}
\frac{1}{k_B T^*}=\left( \frac{3N}{2} - 1 \right)
\frac{\chi}{E}\left< \frac{1}{\chi+ u(\cdot)/N^2} \right> \,,
\end{equation}
where 
\begin{equation} \label{eq:average}
\left< \frac{1}{\chi+u(\cdot)/N^2} \right> = 
\frac{\int _0^1 \cdots \int _0^1 d^{3N} \mb{r}_i 
	\Big[ \chi + u(\cdot)/N^2 \Big]^{3N/2-2}}
{\int _0^1 \cdots \int _0^1 d^{3N} \mb{r}_i 
	\Big[ \chi + u(\cdot)/N^2 \Big]^{3N/2-1}}
\end{equation}
can be interpreted as some sort of average, and still contains 
the integrals over the coordinates to be carried out, to be 
computed in the following. This result coincides with the one 
obtained in \cite{2002VS} for the BG temperature in the 
microcanonical ensemble, i.e., it seems that the physical 
quantities in Tsallis statistics recover the results obtained in 
BG statistics, however with different thermodynamic limits. 
It is important to note that the physical temperature thus 
implicitly contains the information introduced by the Tsallis 
distribution, even though the resulting expression does not 
seem to differ from the original BG case. Already here we 
can see that in the limit of zero gravity, the average 
\eqref{eq:average} becomes 
\begin{equation} \label{eq:av-limit}
\left< \frac{1}{\chi+u(\cdot)/N^2} \right>^{(IG)} = 
\frac{1}{\chi} \,,
\end{equation}
and thus the analog of the ideal gas result for the Tsallis 
temperature is recovered in \eqref{eq:Tphysmicro}.

Using the thermodynamic limit for the gravitational configuration 
integral \eqref{eq:Zcl-limit} in \eqref{eq:SqDerE}, and interchanging 
derivatives and thermodynamic limit significantly simplifies the 
expressions, and the temperature becomes, up to second order, 
\begin{equation}
	\frac{E}{Nk_B T^*} \simeq \frac{3}{2} 
	-\frac{9b_0}{2\chi}-\frac{9}{4\chi^2}(b_1-42b_0^2) 
	\,.
\end{equation}
In the case of zero gravity, $1/\chi = 0$, and thus the result for 
the ideal gas case in Tsallis statistics is obtained. Note that it 
is not identical to the ideal gas in BG statistics, since the 
temperature is the one defined for a Tsallis statistics, and not 
the conventional temperature of BG.

For the calculation of the physical pressure, we take the 
derivative of $S_q$ with respect to the volume, 
\begin{eqnarray} \label{eq:SqDerV}
	\left( \frac{\partial S_q}{\partial V} \right)_{E,N} 
	&=& k_B \left[ \Omega ^{(IG)}\cdot \mathcal{Z}^{(grav)} 
	\right]^{1-q} \\
	&& ~~~ \cdot \left[\frac{\partial}{\partial V} 
	\ln{\Omega ^{(IG)}} + \frac{\chi}{3V}
	\frac{\partial}{\partial\chi} \ln{\mathcal{Z}^{(grav)}} 
	\right] \,. \nonumber
\end{eqnarray}
Substituting this and using the physical temperature 
\eqref{eq:Tphysmicro} in \eqeqref{eq:Pphys}, we obtain 
\begin{eqnarray} \label{eq:Pphysmicro}
	\frac{P^*}{k_B T^*} &=& \frac{1}{3V}\left(\frac{3N}{2} 
	+ 1 \right) \\
	&& ~~~~ + \frac{\chi}{3V}\left(\frac{3N}{2} - 1 \right)
	\left< \frac{1}{\chi+u(\cdot)/N^2} \right> \,, \nonumber
\end{eqnarray}
which is the corresponding Tsallis result to the one obtained 
in the BG statistics in Ref. \cite{2002VS}. 
Again, it is not identical to the BG result, due to the use of the Tsallis 
variables instead of the conventional BG ones. Using 
\eqref{eq:Zcl-limit} in \eqref{eq:SqDerV} and applying the same 
procedure as in the case of the temperature, the equation of state 
is obtained, up to second order, as 
\begin{equation}
	\frac{P^*V}{Nk_B T^*} \simeq 1 - 
	\frac{3b_0}{2\chi} - \frac{3}{4\chi^2} (b_1-42b_0^2) 
	\,. 
\end{equation}
Also for this equation of state, the Tsallis analog of the ideal 
gas is recovered considering \eqref{eq:av-limit}, or equivalently, 
$1/\chi = 0$. This result again reproduces the 
result of \cite{2002VS}, considering the 
different definitions of the $b_i$, and the use of the Tsallis 
variables instead of the conventional BG ones.

As in the canonical ensemble, we will further consider the 
specific heat capacity as defined in \eqeqref{eq:cVcan} for the 
canonical ensemble, derived from \textit{physical} quantities as 
proposed in \cite{2001Abe}. 
It is important to remark that in the microcanonical ensemble 
$(c_{V})_q$ can take positive as well as negative values. In 
\cite{2002VS} it was shown that if fluctuations in the system are 
sufficiently large, this response function becomes negative, implying 
the existence of a phase transition for the self-gravitating gas. 
For the microcanonical ensemble, the expression \eqref{eq:cVcan} 
can be reformulated in more simple terms as 
\begin{equation}\label{eq:SpeHeatqV}
(c_{V})_q = \frac{1}{N}\left( \frac{\partial T^*}
{\partial E} \right)_{V,N}^{-1} \,.
\end{equation}
It is useful, as done in \cite{2002VS}, to define a new function 
in order to calculate the specific heat capacity, 
\begin{eqnarray} \label{eq:g(chi)}
	g(\chi) &=& \frac{\chi}{N}\frac{\partial}{\partial \chi} 
	\ln\left[ \chi^{3N/2-1} \mathcal{Z}^{(grav)} \right] \\
	&=& \frac{\chi}{N} \left( \frac{3N}{2} - 1\right) 
	\left< \frac{1}{\chi+ u(\cdot)/N^2} \right> \,. \nonumber
\end{eqnarray}
Substituting into \eqref{eq:Tphysmicro}, the microcanonical physical 
temperature can be expressed as 
\begin{equation}
T^*= \frac{E}{N k_B g(\chi)} \,,
\end{equation}
and thus 
\begin{equation}
\frac{\partial T^*}{\partial E} = \frac{1}{N k_B} 
\frac{\partial}{\partial E} \left[ \frac{E}{g(\chi)} \right] = 
\frac{1}{N k_B} \left[ \frac{1-\chi \,g'(\chi)/g(\chi)}
{g(\chi)} \right] \,,
\end{equation}
where $g'(\chi)=\partial g(\chi)/\partial \chi$. Therefore, the 
specific heat capacity is given by 
\begin{equation} \label{eq:cvq}
(c_{V})_q = \frac{k_B \, g(\chi)}{1-\chi g'(\chi)/g(\chi)} \,,
\quad \text{or} \quad 
\frac{1}{c_{qV}} = \frac{d}{d \chi}\left[ \frac{\chi}{g(\chi)}
\right] \,.
\end{equation} 
The factor inside the derivative on the right can be written as
\begin{equation} 
\frac{\chi}{g(\chi)} = \frac{2}{(3 - 2/N)} 
\frac{\int _0^1 \cdots \int _0^1 d^{3N} \mb{r}_i 
	\Big[ \chi + u(\cdot)/N^2 \Big]^{3N/2-1}}
{\int _0^1 \cdots \int _0^1 d^{3N} \mb{r}_i 
	\Big[ \chi + u(\cdot)/N^2 \Big]^{3N/2-2}} \,.
\end{equation}
Calculating the derivative with respect to $\chi$, and dropping 
terms of the order of $\mathcal{O} \left(  1/N \right)$, leads to 
\begin{equation} \label{eq:cqV-expansion}
	\frac{1}{c_{qV}} \simeq \frac{2}{3} - 
	\frac{N \left[\left< \left[\frac{1}{\chi+u(\cdot)/N^2} 
	\right]^2 \right> - \left< \frac{1}{\chi+u(\cdot)/N^2} 
	\right>^2  \right]}{\left< \frac{1}{\chi+u(\cdot)/N^2} 
	\right>^2} 
	\,,
\end{equation}
which is the analogous result for the Tsallis statistics as 
found for the BG microcanonical ensemble in \cite{2002VS}. This 
relation can be put in terms of the fluctuation for inverse of 
the physical temperature \eqeqref{eq:Tphysmicro} using $\beta_q$,
\begin{equation} \label{eq:MCEcVqual}
	\frac{1}{(c_{V})_q} = \frac{2}{3} - 
	\left(\frac{\Delta \beta_q}{\beta_q}\right)^2 \,.
\end{equation}
If the fluctuations are large, the specific heat capacity 
can become negative, and the system thus unstable, as was 
shown in \cite{2002VS} as well. This holds in general, i.e., 
not only for weak gravity, since it was obtained from the 
exact expressions before applying the dilute regime.

An explicit expression for the specific heat capacity $c_{qV}$ 
can be found from \eqeqref{eq:cvq}. In order to calculate the 
function $g(\chi)$ from \eqeqref{eq:g(chi)}, we use the 
thermodynamic limit in the dilute regime of $\mathcal{Z}^{(grav)}$ 
given from \eqeqref{eq:Zcl-limit}, to second order resulting in 
\begin{equation}
	\lim_{N\to\infty} g(\chi) \simeq \frac{3}{2} + 
	\frac{9b_0}{2\chi} - \frac{9}{4 \chi^2} (b_1-42b_0^2) 
	\,.
\end{equation}
The specific heat is then obtained from \eqeqref{eq:cvq} by 
derivation with respect to $\chi$, resulting in 
\begin{equation}
\frac{(c_{V})_q}{k_B} = \frac{3}{2} \frac{ 
	\big[ 1 + 6 b_0 \chi^{-1} - 3/2 \, \chi^{-2} (b_1 - 48 b^2_0 
	\big] }{ \big[ 1 + 6 b_0 \chi^{-1} -  9/2 \, \chi^{-2} 
	(b_1 - 42 b^2_0) )\big] }\,.
\end{equation}
Also here we see that in the limit of zero gravity, i.e., 
$1/\chi = 0$, the specific heat also recovers the analog of 
the ideal gas case in the Tsallis statistics. 
From this result, it seems that the heat 
capacity in the microcanonical ensemble does not become 
negative for any value of $\chi$. However, this expression 
only holds in the weak gravitational regime, and thus it 
cannot capture effects that happen in the presence of strong 
gravitational forces, where fluctuations might be large, as 
indicated by \eqref{eq:MCEcVqual}.

\section{Comparisons and Conclusions}
\label{sec:conclusions}
In this work we have applied the framework of Tsallis generalized 
statistics and all its principles to analyze a system of self-
gravitating particles, calculating thermodynamic state and response 
functions of the system  in the appropriate 
thermodynamic limits and investigating their properties. 
After the analyses of canonical and microcanonical ensembles of 
the self-gravitating gas, respectively, we would like to comment 
on a few points concerning comparison and peculiarities of the 
results. 

The application of Tsallis statistics was done in the hopes of 
addressing the question of ensemble equivalence and obtaining 
a consistent description of the self-gravitating gas throughout 
the different ensembles, in particular retaining a traditional 
scaling behaviour for $N$ and $V$. In \cite{2002VS}, ensemble 
equivalence has been achieved, but only by the application of 
a modified "dilute" thermodynamic limit, in which for 
$N,V \to \infty$, the ratio $N/V^{1/3}$ remains constant, 
instead of the ratio $N/V$ as usual. Moreover, closer inspection 
reveals that the limits were indeed applied inconsistently -- 
the constant ratio $N/V^{1/3}$ was required in order to have the 
variable $\eta$ constant, but at the same time terms such as $N/V$ 
have been used to express equations of state, corresponding 
to the conventional thermodynamic limit instead of the dilute one. \\
Considering the outcomes for the thermodynamic equation of 
state in our calculations using the Tsallis statistics, 
it turns out that there is a certain equivalence between 
microcanonical and canonical ensemble, up to a factor 
of $(1-q)$. Leaving aside this factor, we can refer to 
\cite{2002VS}, where an equivalence was shown between the 
two working variables $\chi$ and $\eta$ of the two ensembles, 
and argue that our results equal those of \cite{2002VS}, at 
least in their dependence on $\chi$ and $\eta$. 
We thus achieve equivalence for the equation of state between 
microcanonical and canonical ensemble, up to the additional 
factor of $(1-q)$ which is present in the result for the 
canonical ensemble. The presence of this factor however seems 
like a shortcoming of the theory itself -- both 
ensembles yield results that are of Tsallis nature, by the use 
of Tsallis intensive variables instead of the conventional BG 
ones, but only in the canonical ensemble there is an explicit 
presence of the non-extensivity parameter. We suspect that such 
an explicit dependence should also occur in the microcanonical 
case, and that thus the microcanonical ensemble should be 
re-evaluated. For example, we would expect the entropy in the 
microcanonical ensemble to be formulated purely in terms of 
Tsallis variables, yet a conventional energy $E$ is used to 
calculate the sum over possible microstates. Another way to 
investigate the equivalence of ensembles in a general form is 
to check whether the probability distribution functions of 
microcanonical and canonical ensemble are related by a Laplace 
transformation \cite{1995Grei}, a standard tool of conventional 
thermodynamics. \\
Unfortunately, the microcanonical ensemble is a rarely used and 
somewhat neglected tool in order to describe thermodynamic 
systems, and even more so in the case of Tsallis statistics. 
There are very few instances of the application of the Tsallis 
microcanonical ensemble in the literature, and thus it has rarely 
been put to the test in well-known physical systems. We 
attribute the non-achievement of ensemble equivalence in our 
work to built-in non-equivalences in the ensembles in Tsallis 
statistics, and suspect that perhaps by according all variables 
between the ensembles, and putting them onto the same footing, 
ensemble equivalence might be reached. \\
A different case is the heat capacity. We reproduce the results 
of \cite{2002VS} in both ensembles, up to the factor of $(1-q)$ 
in the canonical case, and they are not equivalent -- not only 
for the explicit $q$-dependence, but also qualitatively. The 
heat capacity stays strictly positive in the canonical ensemble, 
but may become negative in the microcanonical. This fundamental 
difference has also been obtained in \cite{2002VS}, and is 
physically reasonable, since second derivatives of thermodynamic 
state variables are related to the fluctuations of a system, and 
these fluctuations are not necessarily equivalent between 
ensembles. The final results for the heat capacity has not been 
explicitly calculated in \cite{2002VS}, but we do achieve 
agreement qualitatively.

Another question is the limit of zero gravity, i.e., the ideal 
gas case. We have obtained the correct analog of the ideal gas 
contribution in all thermodynamic equations of state and response 
functions calculated, for the case when the gravitational 
interaction is turned off -- analog, because in all the expressions 
the variables are the Tsallis variables, but in the functional form 
they obey the same laws as in the BG case of the ideal gas. 
In the microcanonical ensemble, the zero gravity case corresponds 
to $1/\chi = 0$, and all the results for an ideal gas are 
recovered analogously for this case. In the canonical ensemble, 
zero gravity is mathematically equivalent to the case of $q \to 1$, 
but it is not implied, and should not be confused. Zero gravity 
corresponds to $\eta =0$, so all terms containing $\eta$ are 
eliminated -- which incidentally eliminates most of the explicit 
$q$-dependence in the thermodynamic functions, seemingly resulting 
in the exact BG ideal gas results. However, $q$-dependence is 
retained, namely in the Tsallis intensive variables, which are 
used instead of the BG intensive variables. The zero gravity limit 
of the thermodynamic state functions is thus the $q$-analog of the 
ideal gas in BG statistics, and not the exact BG result. 

Something similar happens with the specific heat in the canonical 
ensemble. The zero gravity case, i.e., $\eta =0$, is mathematically 
equivalent to the case $q \to 1$, but the result in the zero 
gravity case is not equivalent to the ideal gas in BG statistics. 
The difference lies in the definition of the temperature used -- 
our result for the specific heat is valid for a canonical ensemble 
with the Tsallis temperature $T^*$ kept fixed, while the usual BG 
statistics use the conventional notion of temperature. 

In general, it is mathematically unsound to take the limit 
$q\to 1$ in the results for the thermodynamic functions. The 
computation of the final expressions requires the integration 
method described by \cite{1995Prato}, which is valid in the 
region $0<q<1$ only, not including the case $q=1$. Thus taking the 
limit $q\to 1$ is technically not permitted in the results obtained 
after the integral transformations of \cite{1995Prato}, since these 
two mathematical steps do not commute. 

In summary, we have obtained reasonable and physically sound 
results from the application of Tsallis statistics to the case of 
the self-gravitating gas, achieving convergence of important 
thermodynamic functions and state equations under the assumption 
of a thermodynamic limit consistent with the statistical framework 
used. Our results have shown that Tsallis statistics is a viable 
tool for the description of systems with long-range forces, but 
that its application has to be carried out with care, and that 
its results have to be understood in the context of the modified 
statistical framework, and all it implies. We think that the 
question of ensemble equivalence merits more detailed 
investigations, especially on the exact and correct formulation 
of the microcanonical ensemble in Tsallis statistics, where we 
see the potential for consistency between ensembles.

\section*{Acknowledgments}
The authors would like to thank F. Nettel for many fruitful 
discussions.  L. F.  Escamilla-Herrera  thanks  the financial 
support from the Consejo Nacional de Ciencia y Tecnolog\'ia 
(CONACyT, M\'exico). C. Gruber acknowledges support by a Junior 
Fellowship of the Hanse-Wissenschaftskolleg Delmenhorst, and 
from the University of Oldenburg and the Research Training Group 
"Models of Gravity". 
This work was partially supported
by UNAM-DGAPA-PAPIIT, Grant No. 111617.


\bibliography{Tsallis}

\end{document}